\documentclass[aip]{revtex4-1}

\usepackage{graphicx}
\usepackage{float}
\usepackage[version=3]{mhchem}
\usepackage{amssymb,amsmath}

\begin{document}
\title{Off-resonance energy absorption in a linear Paul trap due to mass selective resonant quenching }
\author{I.~Sivarajah}
\affiliation{Department of Physics, University of Connecticut, Storrs, Connecticut 06269, USA}
\author{D.~S.~Goodman}
\affiliation{Department of Physics, University of Connecticut, Storrs, Connecticut 06269, USA}
\author{J. E.~Wells}
\affiliation{Department of Physics, University of Connecticut, Storrs, Connecticut 06269, USA}
\author{F.~A.~Narducci}
\affiliation{Naval Air Systems Command, EO Sensors Division, Bldg 2187, Suite 3190 Patuxent River, Maryland 20670, USA }  
\author{W.~W.~Smith}
\affiliation{Department of Physics, University of Connecticut, Storrs, Connecticut 06269, USA}
\date{\today}
\begin{abstract}
Linear Paul r.f.~ion traps (LPT) are used in many experimental studies such as mass spectrometry, atom-ion collisions and ion-molecule reactions. Mass selective resonant quenching (MSRQ) is implemented in LPT either to identify a charged particle's mass or to remove unwanted ions from a controlled experimental environment. In the latter case, MSRQ can introduce undesired heating to co-trapped ions of different mass, whose secular motion is off resonance with the quenching ac field, which we call off-resonance energy absorption (OREA). We present simulations and experimental evidence that show that the OREA increases exponentially with the number of ions loaded into the trap and with the amplitude of the off-resonance external ac field. 
\end{abstract}
\maketitle
[Copyright 2013 American Institute of Physics.  This article may be downloaded for personal use only.  Any other use requires prior permission of the author and the American Institute of Physics.
The following article appeared in Rev. Sci. Instrum. 84, 113101 (2013) and may be found at http://scitation.aip.org/content/aip/journal/rsi/84/11/10.1063/1.4825352]
\section{INTRODUCTION}
\label{sec:Introduction}

 Ion traps play an important role in mass spectrometry, measuring absolute  mass of stable and radioactive atomic ions \cite{Blaum:06, Jae:02, Makarov:06, Hashimoto:06} to high-accuracy and as an analyzing tool in bio-chemical studies \cite{Coon:05, Husser:03}. Due to its increased ion storage capacity, large trap depths, improved trapping efficiency, and the simplicity of its construction, the linear Paul trap (LPT) is a popular ion trap design. Apart from mass spectrometry, LPTs are widely employed in frequency standard measurements \cite{Blatt:92,Diddams:11}, quantum information applications, \cite{Zoller:95,Everitt:04} and quantum chemistry \cite{Krems:08,Donley:02,Grier:2009, Felix:2011,Rellergert:2011,Schmid:2010,ZipkesPRL:2010,ZipkesNature:2010}. In most of these applications, the trapped ions are cooled to low temperatures to extend storage times, to produce pure quantum states, and to improve spectroscopic resolution. Laser cooling \cite{Raizen:1992,Birkl:1992}, resistive cooling \cite{Wayne:95,Maero:07, Herfurth:06}, or sympathetic cooling by co-trapped pre-cooled ions \cite{Molhave:2000,Blythe:2005,Larson:1986}, chilled buffer gases \cite{DeVoe:2009,Schwarz:2008,Major:1968}, localized ultracold [Bose-Einstein condensates (BEC)] \cite{Schmid:2010,ZipkesPRL:2010,ZipkesNature:2010}, or cold [magneto optical trap (MOT)]\cite{Ravi:11, Siva:12} atomic gases are some of the methods implemented for ion cooling. 

Mass Selective Resonant Quenching (MSRQ) \cite{Hashimoto:06} is a method for selectively driving ions from the trap based on their mass. An rf field is added to the LPT electrodes in addition to the trapping field at a frequency that is resonant with the secular motion of an ion with a particular mass. Typically this method is used to perform mass spectrometry on a sample\cite{Waki:96}, but it has also been used in sympathetic heating spectroscopy in a Paul Trap\cite{Clark:10}.
 
MSRQ can also be used to continuously quench unwanted background ions in a Paul trap \cite{Siva:12,Major:2004,Drakoudis:2006,Sullivan:2011}. In  experiments in a MOT-LPT hybrid trap\cite{Ravi:11, Siva:12}, the MOT is a constant source of molecular and atomic ions via photoassociative or photodissociative reactions with the trapping beams\cite{Gould:1988,Julienne:1991,Tapalian:1994}. In addition, the methods we use to create ions, electron impact ionization (EI) and resonance enhanced multi-photon ionization (REMPI), can also create unwanted ions from background gas atoms or molecules. MSRQ can used to quench these unwanted ions.
 
Applying an extra field to the trap electrodes has the undesirable side effect of significantly perturbing the motion of co-trapped ions, even when their mass dependent secular frequency is far from the applied MSRQ frequency. In this paper, we refer to this effect as off-resonance energy absorption (OREA); in a previous paper we called this effect ac side-effect heating\cite{Goodman:12}. Whenever MSRQ is used on unwanted ions, the experimental sample will undergo OREA, which coherently increases the energy of the individual ions, as well as causing the energy distribution to broaden. This heating of the ion cloud is due to collisions which cause the ions to move incoherently in the rf field.
 
Prior work\cite{Clark:10, Waki:96} on MSRQ and cooling ions focused on ultracold ion clouds or Coulomb crystals, created with laser cooling. In these cases, the ion-ion coupling is strong with the plasma Coulomb coupling parameter (ratio of the Coulomb energy between adjacent particles to the random thermal kinetic energy)\cite{Drewsen:98,Hansen:73}, $\Gamma \sim$ 100) and the MSRQ fields are relatively weak so that few ions are lost. In one case\cite{Waki:96}, the MSRQ frequency is scanned, so ions are on resonance with the field for only a small fraction of the time they are trapped. For the experiments discussed in this work, we investigate higher temperature ion clouds ($T\sim10^3$ K), weak ion-ion coupling ($\Gamma <<$ 1), and strong, constant MSRQ fields.
 
We show that OREA is an effect independent of sympathetic cooling or heating and that it would likely not be seen in an experiment using laser cooling. For example, Baba and Waki \cite{Waki:96} laser cooled \ce{^{24}Mg+} ions; assuming the \ce{^{24}Mg+} ions are saturated, the laser cooling power per ion is $P=\frac{hc}{\lambda \tau} \sim 10^{-10}$ W, where $h$ is Planck's constant, c is the speed of light, $\lambda$ is the wavelength of the cooling photon, and $\tau$ is the natural lifetime of the transition. As will be demonstrated, the typical OREA power per ion is many orders of magnitude smaller, $P \sim 10^{-17}$ W. Therefore, this effect won't be noticeable in laser cooling experiments, but when lower power cooling methods are used, such as sympathetic cooling, OREA can have a comparable effect on the ion energy.

In this paper we present instrumentation to effectively remove unwanted ions via MSRQ. We also give evidence from experiment and simulation (using \textsc{simion} 7.0 software \cite{Manura:2007,Appelhans:2002}) of the dependence of OREA on the applied ac field amplitude and the number of ions trapped.

The paper is organized as follows: In Sec.~\ref{sec:Background} we give a brief background on charged particle trapping as it relates to our instrumentation to implement MSRQ. The results obtained from simulations and experiments are compared in Sec.~\ref{sec:Results} and we conclude in Sec.~\ref{sec:Conclusions}
\section{Background}
\label{sec:Background}
\subsection{Paul trap theory}
\label{subsec:Paul trap theory}

Our hybrid trap experiment for sympathetically cooling \ce{Na+} ions by a cold Na MOT is described in detail elsewhere \cite{Siva:12, Goodman:12, Smith:2003,Smith:2005}. The hybrid trap consists of a Na MOT formed at the geometric center of a LPT, where a trapped cloud of ions is overlapped by the cold MOT atoms, allowing the ions to be sympathetically cooled via elastic scattering or charge exchange collisions.

\begin{figure}[H]
\centering
\includegraphics[width=4 in]{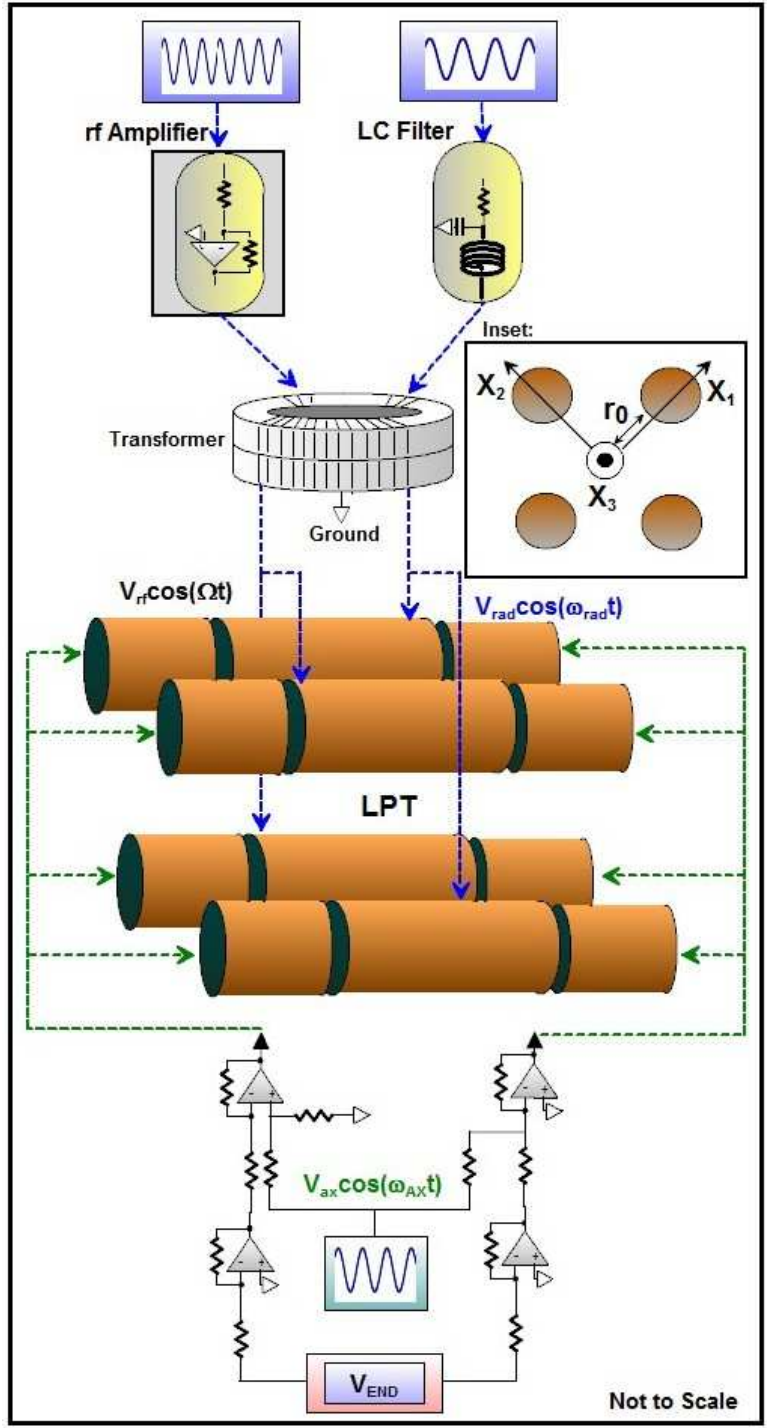}
\caption{(Color online).  Diagram of the LPT showing the electrode configuration  and the rf and dc fields applied for ion trapping. The center of the transformer is grounded so that diagonal rods' fields are precisely $180^{\circ}$ out of phase. Also shown is the  implemention of MSRQ in the radial and the axial directions. Inset: Axial view of Paul trap with Cartesian coordinate system.}
\label{fig:Side_effect1}
\end{figure}

Ion trap theory is well documented  \cite{Raizen:1992, Major:2004,Drewsen:2000,Schuessler:2005,Paul:90, Goodman:12}. Detailed description of this theory, with the assistance of our LPT, is given in our earlier publication \cite{Goodman:12}.  A combination of rf and dc fields  produce a time-dependent electric potential of the approximate form
\begin{equation}
\Phi(x_i,t) \approx   V_{\mathrm{rf}} ~\mathrm{cos}  (\Omega t  ) \frac{x_\mathrm{1}^2-x_\mathrm{2}^2}{r_0^2} +\frac{\eta V_{\mathrm{end}}}{z_0^2} (x_3^2-\frac{x_\mathrm{1}^2+x_\mathrm{2}^2}{2} )
\label{PTpotential}
\end{equation} 
near the center of the LPT (Fig.~\ref{fig:Side_effect1})  when $x_1^2 + x_2^2 \ll r_0^2$. The coordinate system is shown in Fig. \ref{fig:Side_effect1} (Inset) where $x_i$ is the magnitude of a component of the position vector. The driving field has an amplitude  $V_\mathrm{rf}$ and angular frequency $\Omega$. A dc potential $V_\mathrm{end}$ is applied on the end segments with the center segments at dc ground.The distance between the surfaces of the two diagonal electrodes is $2r_0$, the length of the rf segment is $2z_0$ and $\eta$ is a unitless efficiency factor dependent on the geometry of the specific LPT. The experimental values of these parameters for our LPT can be found in Table~\ref{LPT}.

\begin{table}[H]
\centering
\caption{LPT settings $\&$ dimensions.}
\begin{ruledtabular}
\begin{tabular}{c c c c c c}
$V_\mathrm{rf}$ & $\Omega/2\pi$ & $V_\mathrm{end}$ & $r_0$ & $z_0$ & $\eta$ \\
\hline
36 V & 729 KHz & 35 V  & 9.5 mm & 24 mm & 0.1 \\
\end{tabular}
\end{ruledtabular}
\label{LPT}
\end{table}

The motion of the ion in an LPT, as described \cite{Goodman:12},  is a superposition of a fast micromotion at the driving frequency $\Omega$ and a slow secular motion $\omega$ due to the trap's pseudopotential \cite{Major:2004}. The frequencies of the ion's secular motion in the radial ($x_1, x_2$) and axial ($x_3$) degrees of freedom, consistent with the coordinate system shown in Fig.~\ref{fig:Side_effect1}, are given by

\begin{equation}
	\omega_\mathrm{rad} \approx \frac{\Omega}{2} \sqrt{a_{1}+\frac{q_{1}^2}{2}} 
	\label{SecFreq1}
\end{equation}
and
\begin{equation}
	\omega_\mathrm{ax} \approx \frac{\Omega}{2} \sqrt{a_3} .
	\label{SecFreq2}
\end{equation}
where $a_{i}$ and $q_{i}$ are LPT stability parameters, which are a function of the ion's mass and the LPT settings \cite{Goodman:12}. Equations (2) and (3) are approximate expressions for $(a,q) \ll 1$, with no shifts from MSRQ included.\cite{Major:2004}

The pseudopotential trap depths are $D_\mathrm{Radial} = \frac{e q_1 V_{\mathrm{rf}}}{4}-\frac{e \eta V_\mathrm{end} r_0^2}{2 z_0^2}$ and $D_\mathrm{Axial} =\eta eV_{\mathrm{end}}$ in the radial and axial directions, respectively \cite{Wineland:92}. When an external ac field is applied to the trapped ions at their radial or axial secular frequency, the ions' energy is resonantly driven above the trap depth $D_\mathrm{Radial}$ or $D_\mathrm{Axial}$.  Since the secular frequency of the ions trapped within the LPT is mass dependent, ions with specific masses can be selectively quenched from the trap using this technique.

The electronic setup for our experimental implemention of MSRQ is shown in Fig.~\ref{fig:Side_effect1}.  The additional radial quenching ac field $V_\mathrm{rad} \cos{(\omega_\mathrm{rad}t)}$ was coupled with the rf driving field via a ferrite toroidal transformer.  An LC filter was used to minimize any feedback to the $V_\mathrm{rad}$ function generator from the large amplitudes associated with the driving rf field. Similarly, using operational amplifiers, the additional axial quenching ac field $V_\mathrm{ax} \cos{(\omega_\mathrm{ax}t)}$ was added to the end segment dc voltage. For the results presented in this paper, $V_\mathrm{rad}$ and $V_\mathrm{ax}$ are applied individually and not as a combination. Applying them separately allows us to distinguish the effect of each field individually.

\subsection{Paul trap experiment}
\label{subsec:LPT experiment}

We loaded \ce{Na+} ions in the LPT by the REMPI method from Na vapor within the vacuum chamber \cite{Compton:1980,Siva:12} while \ce{Ca+} ions are loaded via EI of Ca vapor. Once ions are loaded for a duration of $t_\mathrm{Load}$, the ions are trapped for a duration $t_\mathrm{Trap}$, after which the the remaining ions are extracted (for a duration $t_\mathrm{Extract}$) by applying a dipole field to the LPT's end segments. A Channeltron electron multiplier (CEM) and a preamplifier collect and detect the ions, producing a current signal whose amplitude is proportional to the number of ions extracted. Using a crude calibration method\cite{Siva:12}, we estimate the order of magnitude of the number of ions typically loaded to be $\sim 10^3 - 10^4$.  We estimate the initial temperature of our ions to be $\sim 10^3$ K (see Sec.~\ref{subsec:EXPResults}). Using the temperature and the theoretical spring constant for the LPT\cite{Lee:2013}, we estimate the typical initial ion cloud mean volume to be $\sim 0.01$ cm$^3$, leading to a dimensionless plasma coupling parameter for 1000 ions \cite {Drewsen:98,Hansen:73} $\Gamma \sim$ 0.001. 
 
\subsection{Associative Ionization and Impact Ionization}
\label{subsec:AI}

Our group's primary research interest involves experiments which use a hybrid ion-neutral Na MOT trap apparatus. The Na MOT acts as an ion source producing \ce{Na2+} and \ce{Na+} ions~\cite{Gould:1988,Julienne:1991}. Excited Na(3p) atoms can form \ce{Na2+} molecular ions via binary associative ionization (AI) reactions. These AI-produced \ce{Na2+} ions can subsequently photodissociate into \ce{Na+} ions and Na atoms via collisions with excited (3p) atoms or by resonant photodissociation with 589 nm photons from the Na MOT beams \cite{Tapalian:1994}.

Since our group's experiments primarily study the interaction between \ce{Na+}(or sometimes \ce{Ca+}) ions with an Na MOT, the additional \ce{Na2+} ions produced from the MOT via AI during $t_\mathrm{Trap}$ (which are contaminants in this experiment) have to be removed. Therefore, we employed MSRQ to remove all \ce{Na2+} produced by AI during $t_\mathrm{Trap}$. Since MSRQ can lead to OREA, which is unfavorable, we have investigated the effects of OREA in detail.

Although not considered in the results of this manuscript, it is worth noting that when we use EI to produce \ce{Ca+} ions from a gas mixture containing \ce{Ca} and \ce{Na} atoms, \ce{Na} atoms also get ionized because they have a lower ionization energy than \ce{Ca}. Since we are only interested in trapping a single ionic species at a time and we study their interaction with the \ce{Na} MOT produced from the \ce{Na} gas, trapped \ce{Ca+} ions produced via this method experience OREA due to the necessary quenching of \ce{Na+} ions produced as a byproduct of EI. 

\subsection{Simulations}
\label{subsec:Simulations}

Simulations for our hybrid trap were done using \textsc{simion} software. Using  \textsc{simion 7.0} we programmed custom simulations of the ion trajectories within the dc and time dependent fields produced by the LPT electrodes, as well as the Coulomb repulsion forces between the ions. \textsc{simion} contains a custom user-programming environment, which allows us to control initial conditions and to monitor the ion's time-dependent position and kinetic energy. Our specific program is described in more detail elsewhere \cite{Goodman:12}.

For simulations of a single trapped ion, the ion's position is initialized at the geometric center of the trap with a specified initial velocity at an azimuthal angle and a polar angle of $45^{\circ}$ relative to the trap axis. In simulations of multiple trapped ions, the ions' have a given initial speed in an  isotropically distributed direction.
 
Simulating initially colder ions speeds up computation time, therefore the simulated initial ion energy was set to $1\times 10^{-8}$ eV (T $\approx $ 0.1 mK), which is much lower than the typical experimental ion temperature.  However all the LPT settings were initialized to match the experimental values as described in Table. \ref{LPT}.

\section{RESULTS}
\label{sec:Results}

OREA was investigated  via simulations and experiments for MSRQ applied both on-resonance with the ion secular frequency and off-resonance with the ion secular motion. For example, on-resonance implementation was studied to find out the optimal  conditions in which to apply MSRQ for experiments involving the removal of AI produced \ce{Na2+} ions. The MSRQ field for removing  \ce{Na2+} ions is off-resonance with the secular frequency of \ce{Na+} ions. Therefore, studying the trap loss of \ce{Na+} ions under the influence of an  MSRQ field at the \ce{Na2+} secular frequency allowed us to study a realistic case of OREA.

\subsection{On-Resonance Results}
\label{subsec:SIMResults}

The motion of a single ion was simulated to determine the expected  $\omega_\mathrm{rad}$ and $\omega_\mathrm{ax}$ of each species in our trap. The frequency of the MSRQ field parameter was scanned until the frequency of the field came into resonance with the ion's secular motion, ultimately ejecting the ion from the trap.

\begin{figure}[H]
\centering
\includegraphics[width=5 in]{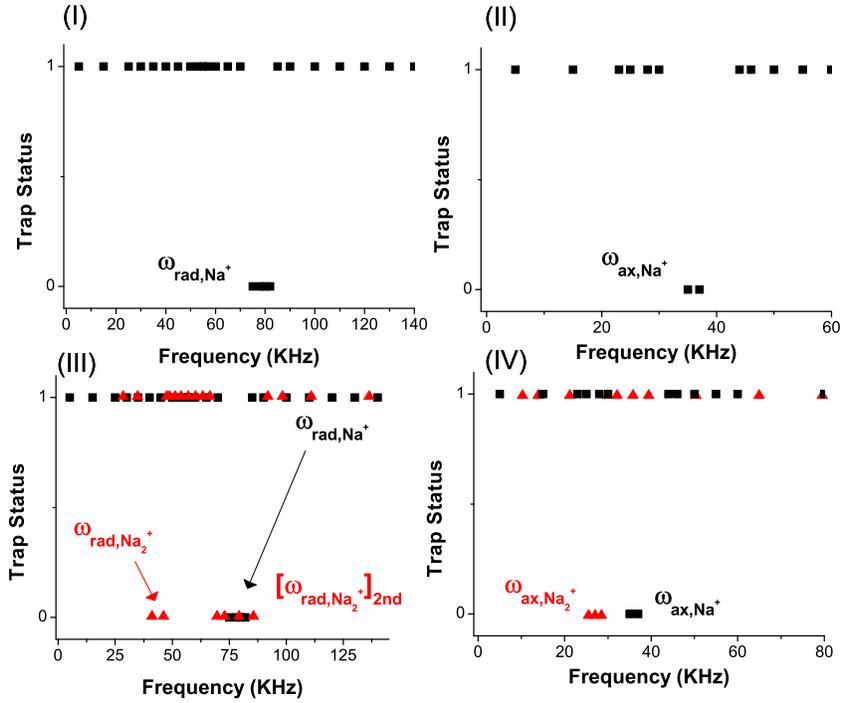}
\caption{(Color online). Simulation results of MSRQ. Trap status is 1 if the ion is still trapped in the LPT and 0 when the ion is ejected. (I) Simulation results showing MSRQ of \ce{Na+} via radial quenching. (II) Simulation results for MSRQ of \ce{Na+} via axial quenching. (III) Simulation results showing MSRQ of \ce{Na2+} (red triangles) and \ce{Na+} (black squares) via radial quenching. Black squares are copied from (I). (IV) Simulation results for MSRQ of \ce{Na2+} (red triangles) and \ce{Na+} [black squares, copied from (II)] via axial quenching. As the frequency is varied, the transition between trapping and ejection is quite abrupt, as shown. All frequencies are circular frequencies (kHz), not angular frequencies.}
\label{fig:Simion_quench}
\end{figure}

The results of the radial and axial frequency sweep simulations are shown in Fig.~\ref{fig:Simion_quench} for \ce{Na2+} or \ce{Na+} ions. When the ion remains in the trap, the trap status is denoted by 1 and when it is ejected the trap status is denoted by a 0. In our simulations, we set the sinusoidal amplitudes to  $V_\mathrm{rad,\ce{Na2+}}$= 0.75 V, $V_\mathrm{ax,\ce{Na2+}}$= 18 V, $V_\mathrm{rad,\ce{Na+}}$ = 0.5 V and $V_\mathrm{ax,\ce{Na+}}$ = 16 V and examined their effect on the motion of the ions. These amplitudes were sufficient to eject the ions from the LPT. The second harmonics of the radial secular frequencies for \ce{Na2+} or \ce{Na+} ions are also seen to quench the trapped ions as seen in Fig.~\ref{fig:Simion_quench}. Consistent with other groups' results \cite{chu98,Drakoudis:2006,Major:2004}, the resonance dips are broader in the second harmonics. The time dependence of the position components of a single simulated ion was found to be approximately sinusoidal.

\begin{figure}[H]
\centering
\includegraphics[width=5 in]{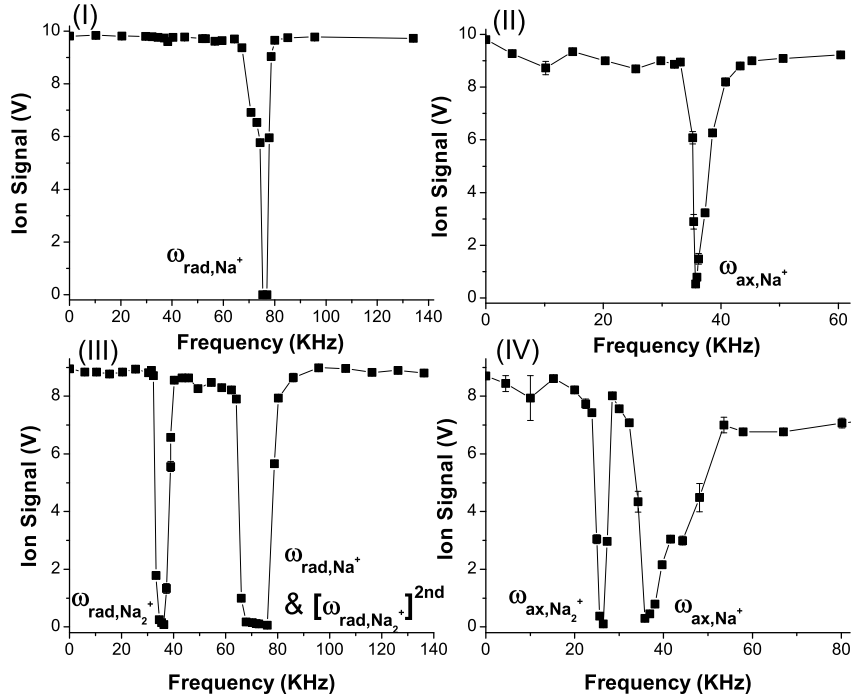}
\caption{(Color online). Experimental results for MSRQ. (I). Experimental results showing MSRQ of \ce{Na+} produced via REMPI by radial quenching. (II) Experimental results for MSRQ of \ce{Na+} produced via REMPI by axial  quenching. (III) Experimental results for MSRQ of \ce{Na2+} and \ce{Na+} ions (produced by AI from the MOT/photodissociation) via radial quenching. (IV) Experimental results showing MSRQ of \ce{Na2+} and \ce{Na+} ions (produced by AI/photodissociation) via axial quenching. Since \ce{Na2+} ions in our experiments are only produced by AI, which dissociates into \ce{Na+} ions, MSRQ was not performed exclusively on a pure sample of \ce{Na2+} ions. The connecting lines are shown to guide the eye. All frequencies are circular frequencies (kHz), not angular frequencies.}
\label{fig:EXPquench}
\end{figure}

Experimentally, $\omega_\mathrm{rad}$ and $\omega_\mathrm{ax}$ were determined by scanning the frequency of the MSRQ field and measuring the number of ions remaining.  The scan was performed on \ce{Na+} ions loaded into the LPT via REMPI from the background  gas or \ce{Na2+} and \ce{Na+} ions produced from the Na MOT by AI and photodissociation. The results of these scans are shown in Fig.~\ref{fig:EXPquench}. The  $\omega_\mathrm{rad,\ce{Na+}}$,  $\omega_\mathrm{ax,\ce{Na+}}$,  $\omega_\mathrm{rad,\ce{Na2+}}$ and  $\omega_\mathrm{ax,\ce{Na2+}}$ found experimentally show very good agreement with the values obtained from simulations,  Eq.~\ref{SecFreq1} and Eq.~\ref{SecFreq2}.

\begin{figure}[H]
\centering
\includegraphics[width=4.7 in]{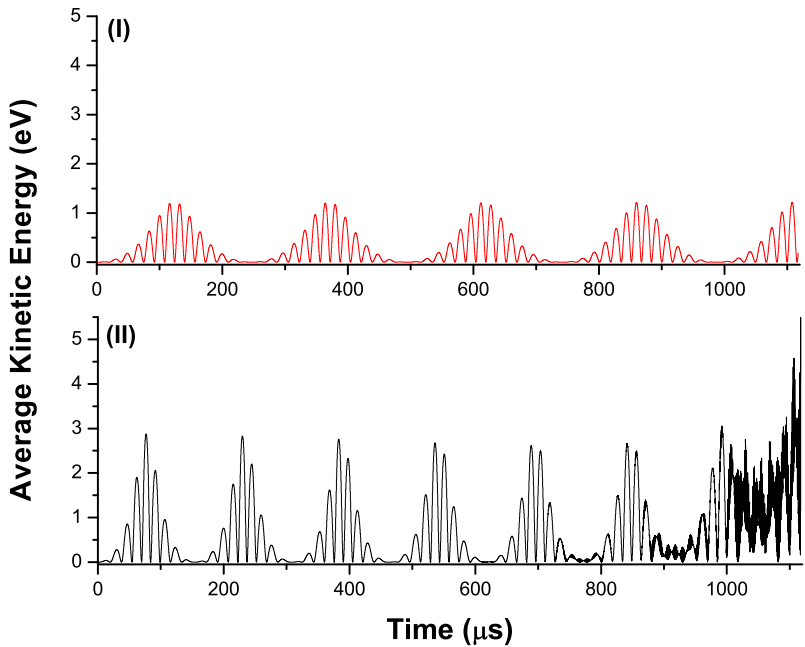}
\caption{(Color online). Simulated time evolution of a single \ce{Na2+} ion's mean kinetic energy due to MSRQ in the axial direction. The rapid oscillations reflect the average of the ion's axial and radial secular frequencies as shown in the previous figure. The quenching field amplitudes are set at (I- red online) $V_\mathrm{ax}$= 9 V and (II - black online) $V_\mathrm{ax}$ = 18 V. Ions are ejected more quickly when the amplitude is increased. The ion in (II) gets ejected at a somewhat later time than shown in the plot.  }
\label{fig:Quenchtime}
\end{figure}

\begin{figure}[H]
\centering
\includegraphics[width=4.7 in]{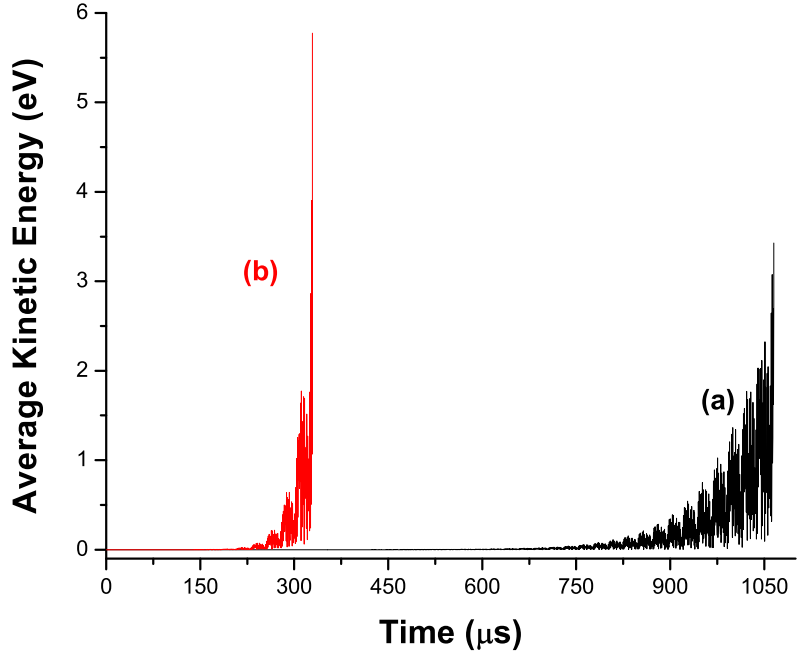}
\caption{(Color online). Simulated time evolution of a single \ce{Na2+} ion's mean kinetic energy due to MSRQ in the radial direction. The quenching field amplitudes are set at (a, black online) $V_\mathrm{rad}$ = 0.75 V and (b, red online) $V_\mathrm{rad}$= 1.5 V. Ions are ejected more quickly when the amplitude is increased. The simulations end when the ion's energy exceeds the radial trap depth and the ion is lost from the trap.  }
\label{fig:Quenchtime2}
\end{figure}

The quenching rate from MSRQ has to be greater than the production rate due to AI, to completely eliminate the unwanted ion population. Therefore, the ion ejection efficiency was tested at different amplitudes of the applied quenching field. The lifetime of a single \ce{Na2+} ion in our trap was measured as a function of the quenching field amplitude in our simulations. The ion's time averaged kinetic energy (averaged over the rf micromotion period) was recorded. The ion's kinetic energy is plotted against time, until the ion was ejected from the LPT in Fig.~\ref{fig:Quenchtime} (axially) and Fig.~\ref{fig:Quenchtime2} (radially). The ion's kinetic energy oscillates due to the secular motion, which can be seen in the figures. Ejection is defined to occur at the time when the simulated kinetic energy grows to exceed the trap depth. For example, in Fig.~\ref{fig:Quenchtime}(I) the ion is ejected at a time beyond the domain of the plot, but Fig.~\ref{fig:Quenchtime}(II) shows some rapid heating after a few secular oscillations around 1000 $\mu$s and then ejection just after 1100 $\mu$s.

\begin{figure}[H]
\centering
\includegraphics[width=5 in]{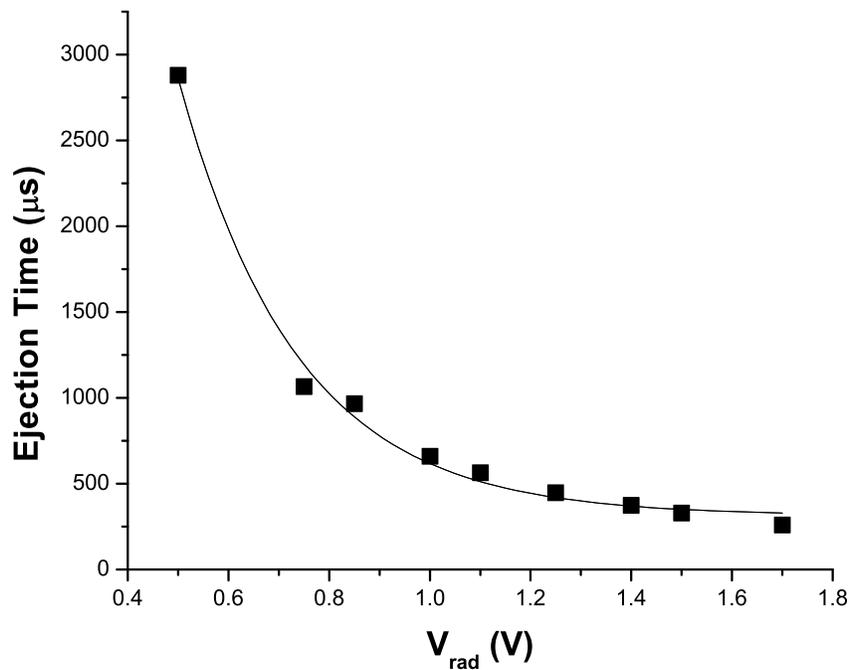}
\caption{(color online). Simulation results showing ejection time for a single \ce{Na2+} ion as a function of amplitude  $V_\mathrm{rad}$. The fit is to a decreasing exponential.}
\label{fig:Amp_time}
\end{figure}

A plot of the ejection time as a function of $V_\mathrm{rad}$ (Fig.~\ref{fig:Amp_time}) shows how ions can be effectively ejected with increasing $V_\mathrm{rad}$. The ejection time is seen to have a decreasing exponential dependence on the applied amplitude $V_\mathrm{rad}$.

\begin{figure}[H]
\centering
\includegraphics[width=5 in]{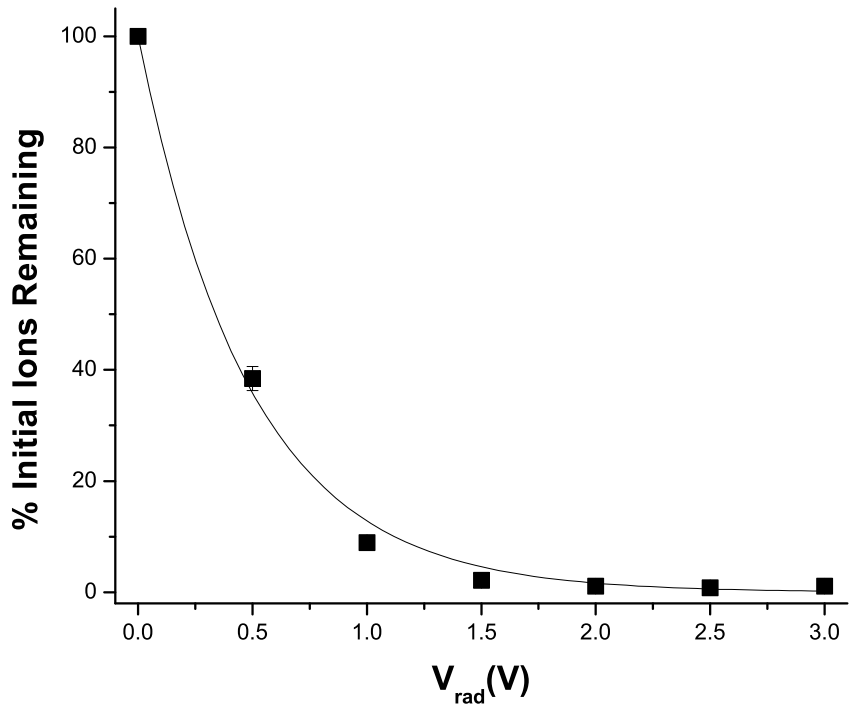}
\caption{(color online). Experimental results showing trap loss from the LPT after $t_\mathrm{Trap}$ = 0.5 s, of \ce{Na2+} ions via MSRQ as a function of amplitude $V_\mathrm{rad}$. The fit is to a decreasing exponential. Where not shown, error bars are smaller than the size of the points. }
\label{fig:QuenchAmp}
\end{figure}

We measured the field amplitude dependence by increasing  $V_\mathrm{rad}$ while monitoring the \ce{Na2+} ions remaining after $t_\mathrm{trap}$ = 0.5 s. As seen in Fig.~\ref{fig:QuenchAmp}, ions are ejected rapidly with increasing $V_\mathrm{rad}$ and show an exponentially decreasing ion population as a function of increasing MSRQ field amplitude, as would be expected from the simulation results.

Due to the fact that our LPT, in practice uses $D_\mathrm{Radial} < D_\mathrm{Axial}$, the radial quenching is faster and requires lower ac amplitudes. Also, since the radial distance from the center of the trapping region to the rf segments is much shorter than the distance to the end segments ($ r_0 \ll z_0$) for our LPT, ions driven in the radial direction are more easily lost from the trap as compared to the axial direction. Experimentally, we found that in order to remove ions at a given rate, we needed to use a $V_\mathrm{ax} \gg V_\mathrm{rad}$. 
 
\subsection{Off-Resonance Results}
\label{subsec:EXPResults}

\begin{figure}[H]
\centering
\includegraphics[width=5 in]{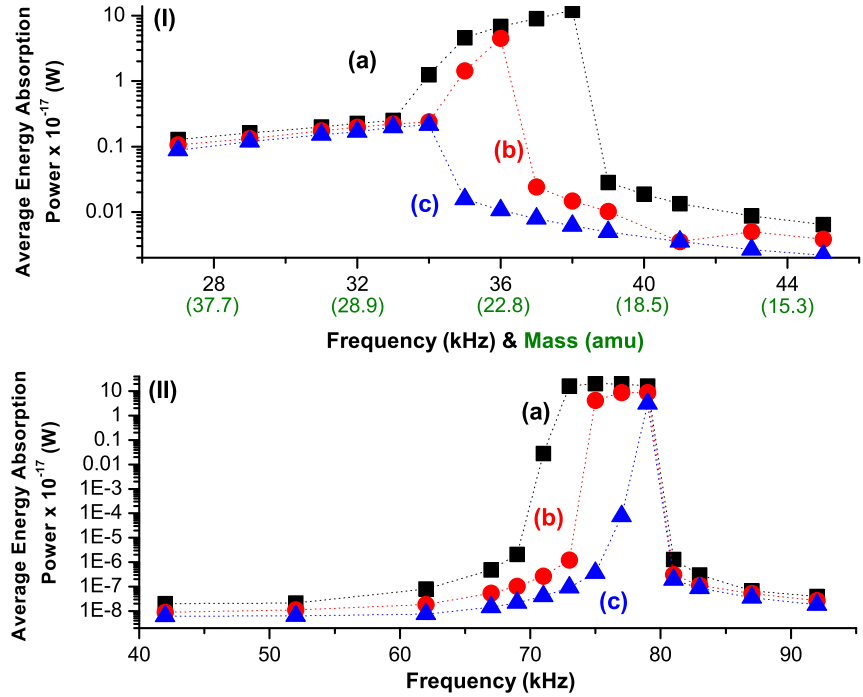}
\caption{(Color online).  Simulation results showing the average energy absorption rate of a single \ce{Na+} ion as a function of (I) $\omega_\mathrm{ax}/2\pi$ for (a, squares, black online) $V_\mathrm{ax}$ = 15 V, (b, dots, red online) $V_\mathrm{ax}$ = 12 V, and (c, triangles, blue online) $V_\mathrm{ax}$ = 9 V. and (II) as a function of $\omega_\mathrm{rad}/2\pi$ for (a, squares, black online) $V_\mathrm{rad}$ = 2.1 V, (b, dots, red online) $V_\mathrm{rad}$ = 1.5 V, and (c, triangles, blue online) $V_\mathrm{rad}$ = 1 V.  In the horizontal axis of (I) we included the mass of an ion (in amu units, in parenthesis) that would be resonant at the specified frequency according to  Eq.~\ref{SecFreq2}. The lines connecting the points are to guide the eye.}
\label{fig:Freq_Dep}
\end{figure}

We investigated in finer detail the OREA due to an applied MSRQ field over a range of frequencies. Simulations were carried out at three different axial (radial) MSRQ amplitudes, measuring the average energy absorption rate of a single \ce{Na+} ion, while the off-resonance frequency was changed from 25 kHz to 45 kHz axially (42 kHz to 92 kHz radially), passing through the \ce{Na+} resonance. The results of the simulation are shown in Fig.~\ref{fig:Freq_Dep}, which plots the energy absorption power of a \ce{Na+} ion versus both the MSRQ frequency and the mass that would be resonant with the MSRQ field at that frequency (according to Eq~\ref{SecFreq2}). For example, when resonantly quenching \ce{CO+} that has a mass of 28 amu, according to Fig.~\ref{fig:Freq_Dep} (I), the OREA by a single \ce{Na+} ion would have a power of at least 2$\times 10^{-18}$ W.

As expected, at low MSRQ amplitudes, the average heating power is approximately at a maximum when the MSRQ frequency comes into resonance with the ion's secular frequency. When the amplitude of the MSRQ is increased, the resonance widens. The asymmetry in the resonance curve is similar to that of an anharmonic driven oscillator. The asymmetry in the resonance peak and its dependence on the applied MSRQ amplitude arises from both the perturbation caused by the MSRQ field itself and the fact that the actual trap electrode's do not create a perfect quadrupole field\cite{Drakoudis:2006}.

To simulate the ejection of one ion species while continuing to trap a \emph{different} ion species, simulations were performed with a single trapped \ce{Na+} ion while MSRQ was applied at either the measured radial or axial fundamental secular frequency of \ce{Na2+} ions (which is clearly off-resonance from \ce{Na+} ions).

\begin{figure}[H]
\centering
\includegraphics[width=5 in]{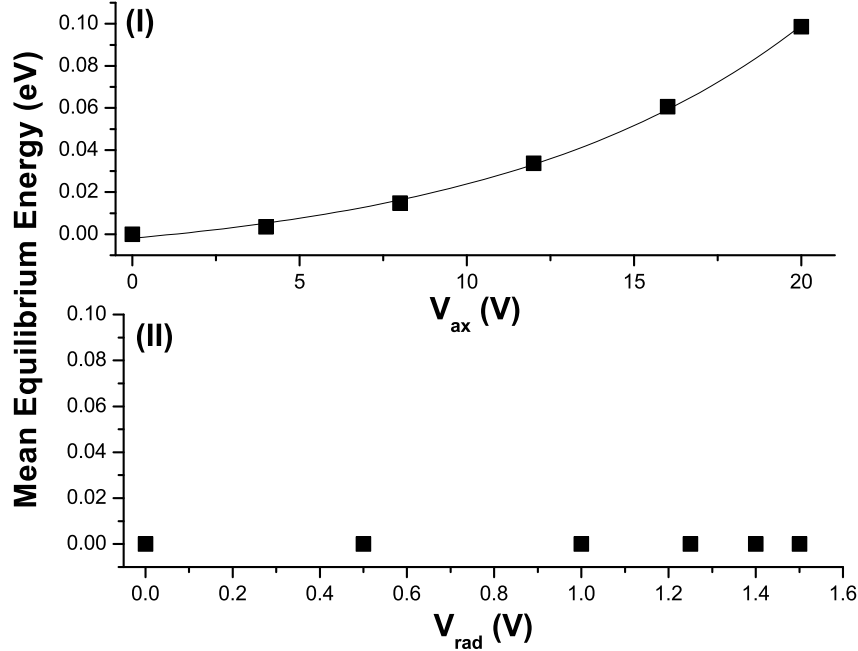}
\caption{(color online). (I) Simulation results showing mean equilibrium energy of a single \ce{Na+} ion subject to axial MSRQ resonant with $\omega_\mathrm{ax, \ce{Na2+}}$  after $t_\mathrm{Trap}$ = 10 ms plotted against increasing $V_\mathrm{ax}$. The fit is to an exponential. (II) Simulation results showing mean equilibrium energy of a single \ce{Na+} ion subject to radial MSRQ resonant with $\omega_\mathrm{rad, \ce{Na2+}}$ after $t_\mathrm{Trap}$ = 10 ms plotted against increasing $V_\mathrm{rad}$ also shows no increase on this scale. OREA is much greater when axial MSRQ is employed.}
\label{fig:SIM_offres}
\end{figure}

\begin{figure}[H]
\centering
\includegraphics[width=5 in]{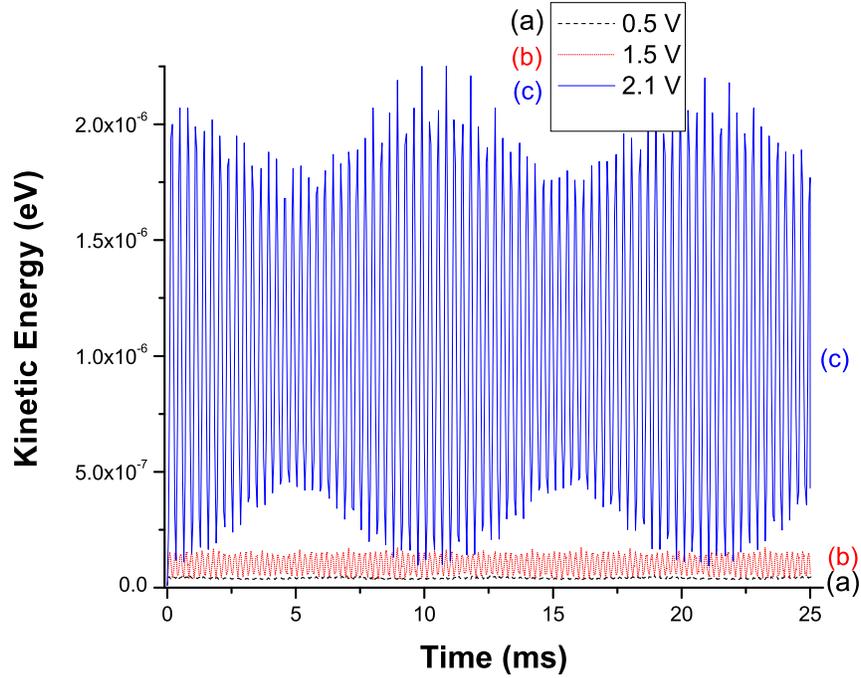}
\caption{(Color online). Simulation results showing the kinetic energy (time averaged over an r.f. micromotion period) of a single \ce{Na+} with MSRQ set for $\omega_\mathrm{rad, \ce{Na2+}}$ as a function of time at three different amplitudes (see inset). The increase in MSRQ field amplitude ($V_\mathrm{rad}$) increases the trapped ions kinetic energy oscillations due to OREA.}
\label{fig:Avg_energy}
\end{figure}

The dependence of OREA (leading to trap loss) of \ce{Na+} ions on the amplitude of the \ce{Na2+} MSRQ field was tested.  We measured the change in average  kinetic energy of a single \ce{Na+} as a function of $V_\mathrm{rad}$ or $V_\mathrm{ax}$ applied at the \ce{Na2+} resonant frequency. We found that after $t_\mathrm{Trap} \approx$ 10 ms the ion reached a quasi-thermal equilibrium. Simulations showed that the mean equilibrium kinetic energy of the ion  increased exponentially with increasing amplitudes (as seen in Fig.~\ref{fig:SIM_offres}). For a single \ce{Na+} ion driven by an off-resonance MSRQ field, higher radial MSRQ amplitudes cause larger kinetic energy oscillations. Our results also show that radial quenching causes less OREA than axial quenching at settings associated with equivalent quenching rates for our trap geometry. Simulation results of a single \ce{Na+} ion energy's time dependence while experiencing OREA due to MSRQ of \ce{Na2+} ions are shown in Fig.~\ref{fig:Avg_energy}. 

\begin{figure}[H]
\centering
\includegraphics[width=5 in]{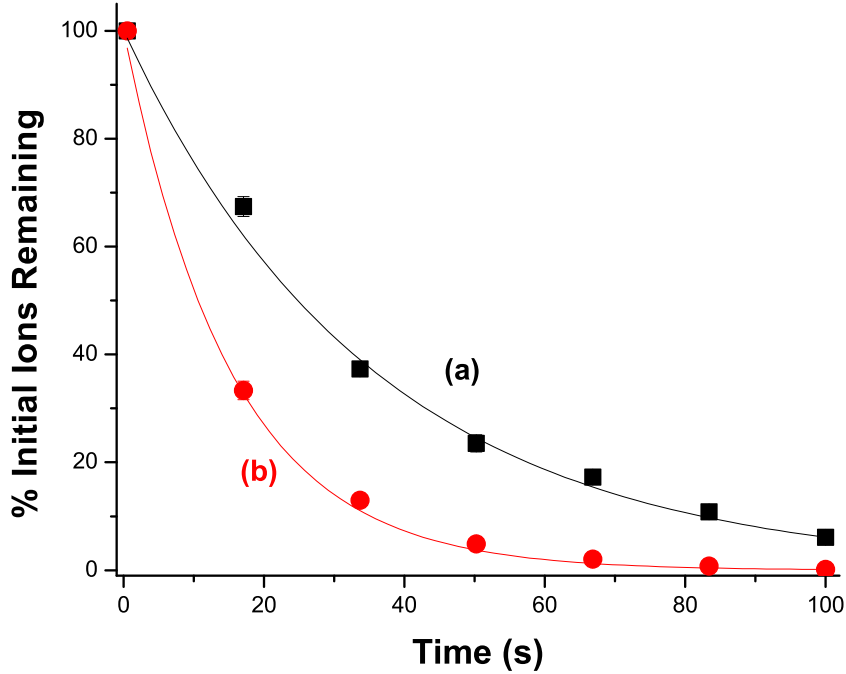}
\caption{(Color online). Experimental results showing OREA. (a) Trap loss of \ce{Na+} ions from the LPT with no MSRQ applied (black squares). (b) Increased trap loss rate of \ce{Na+} ions from the LPT with MSRQ set for $\omega_\mathrm{rad, \ce{Na2+}}$ (off resonance for \ce{Na+} ions - red circles). This  shows OREA facilitates trap loss. Fits are to decreasing exponentials, where the time constant for (a) is $\approx$ 35 s and the time constant for (b) is $\approx$ 15 s. Error bars are smaller than the size of the points.}
\label{fig:Heating}
\end{figure}

Experimentally, one way in which OREA was observed was by monitoring trap loss from the LPT as a function of $t_\mathrm{Trap}$ at a constant trap depth. After loading a sample of \ce{Na+} ions in the LPT, the time dependence of the number of trapped ions was recorded. When MSRQ is introduced to the LPT, at $\omega_\mathrm{rad,\ce{Na2+}}$, the number of \ce{Na+} ions decreases faster from the LPT [Fig.~\ref{fig:Heating} (b)] than without it  [Fig.~\ref{fig:Heating} (a)].

To experimentally test the OREA effects on the energy distribution of the trapped \ce{Na+} ions, the change in trap depth technique described in our previous publication~\cite{Siva:12} was used. In this method, a sample of initial \ce{Na+} ions is loaded and trapped for a fixed $t_\mathrm{Trap}$ = 10 s, then the radial trap depth ($D_\mathrm{Radial}$) is suddenly lowered momentarily (for 10 ms) right before extraction. When $D_\mathrm{Radial}$  is lowered, ions whose energy are above the minimum  $D_\mathrm{Radial}$ will evaporate from the trap. \ce{Na+} ions were loaded into the LPT, while MSRQ was applied at the first harmonic of \ce{Na2+} with $V_\mathrm{rad}$  amplitudes of 0V, 1V and 2V (off resonant to \ce{Na+)}.

\begin{figure}[H]
\centering
\includegraphics[width=5 in]{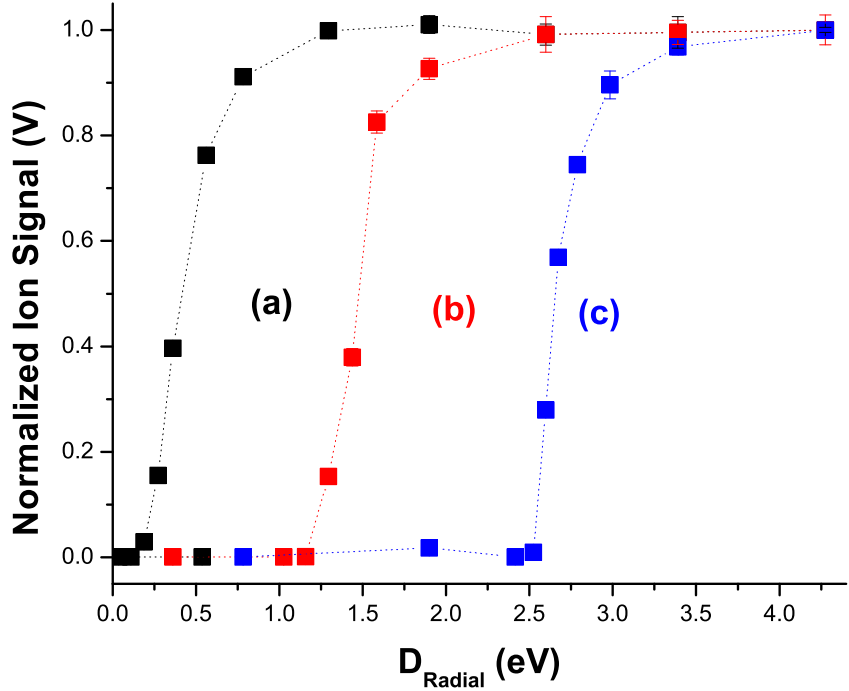}
\caption{(Color online).  Fraction of ions trapped \ce{Na+} ions remaining in the trap as a function of a sudden change in final trap depth ($D_\mathrm{Radial}$), just before extraction after a fixed $t_\mathrm{Trap}$ = 10 s, starting with an initial $D_\mathrm{Radial}$ = 4.5 eV for the 3 cases. $V_\mathrm{rad}$ was set for (a) 0 V (black points), (b) 1 V (red points) and (c) 2 V (blue points). Ions are shown to have greater mean energy ($D_\mathrm{Radial}$ value half way up the cliff) at high $V_\mathrm{rad}$ due to OREA at a secular frequency tuned to $\omega_\mathrm{rad, \ce{Na2+}}$ (i.e. off resonance with the \ce{Na+} ions). The dotted lines are to guide the eye. When not visible, error bars are smaller than the size of the points.}
\label{fig:Qdrop}
\end{figure}

Our results show that changing $V_\mathrm{rad}$ changes the energy distribution. For $V_\mathrm{rad} = 2$ V the \ce{Na+} ion signal reaches 50$\%$ of its initial value at $D_\mathrm{Radial}\approx 2.75$ eV, as compared to smaller $D_\mathrm{Radial}\approx 1.5$ eV for 1 V amplitude and $D_\mathrm{Radial}\approx 0.4$ eV when there is no MSRQ (see Fig.~\ref{fig:Qdrop}). Making the approximation that the mean energy per ion is the post-drop trap depth when 50$\%$ of the population is lost, we can determine the temperature of the ion cloud, e.g., in the absence of MSRQ the ions are $\sim 10^3$ K.

We believe that the curves in Fig.~\ref{fig:Qdrop} do not go to zero at the same ($D_\mathrm{Radial}$) because OREA promotes the coldest ions to a non-zero minimum energy.  Furthermore, as the MSRQ field increases, the slopes of the curves become steeper, indicating that the energy distribution becomes narrower. The distribution narrows because the minimum energy of the ions increases, but, since the initial trap depth does not change, the top of the distribution is truncated. The effect of a truncated distribution has been observed in other unrelated experiments\cite{Zeppenfeld:2012}.

\begin{figure}[H]
\centering
\includegraphics[width=5 in]{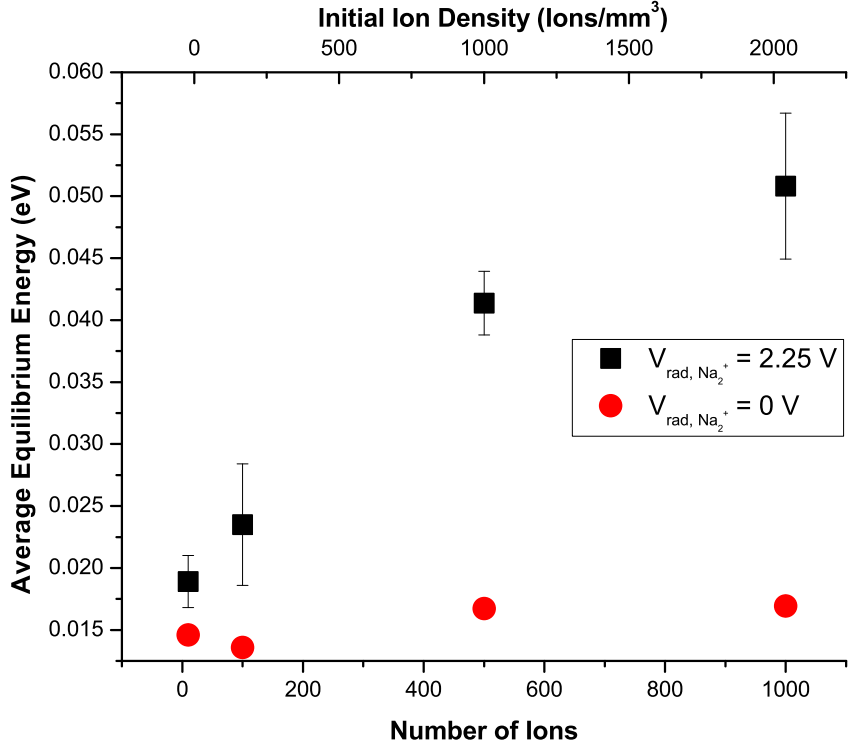}
\caption{(Color online). Simulation results showing the dependence of OREA on initial ion number (bottom x axis) and on initial ion density (top x axis) for fixed $V_\mathrm{rad,\ce{Na2+}}$ of 2.25 V and  0 V (see inset). Without MSRQ the ion-ion rf heating is minimal and the mean energy remains fairly constant. With MSRQ, the OREA equilibrium energy increases approximately linearly with increasing initial ion number and initial ion density. These simulations are performed for weak coupling\cite{Drewsen:98,Hansen:73}, $\Gamma \sim 0.01.$}
\label{fig:SIM_ion_number}
\end{figure}

The dependence of OREA on the number of \ce{Na+} ions and the initial ion density was also determined by simulations. We tested this with MSRQ resonant with the \ce{Na2+} secular frequency. We monitored the simulated ion cloud's mean kinetic energy and the standard deviation of the energy fluctuations after the cloud reached quasi-thermal equilibrium at $t_\mathrm{Trap} \approx$ 3 ms. When the initial number of ions are increased we found that the mean kinetic energy increased due to OREA as seen in Fig.~\ref{fig:SIM_ion_number}. However, reducing the MSRQ amplitude weakens the OREA, which suggests that it is possible to find conditions where MSRQ can be implemented radially with minimal OREA, provided the desired quenching rate can still be achieved.

\begin{figure}[H]
\centering
\includegraphics[width=5 in]{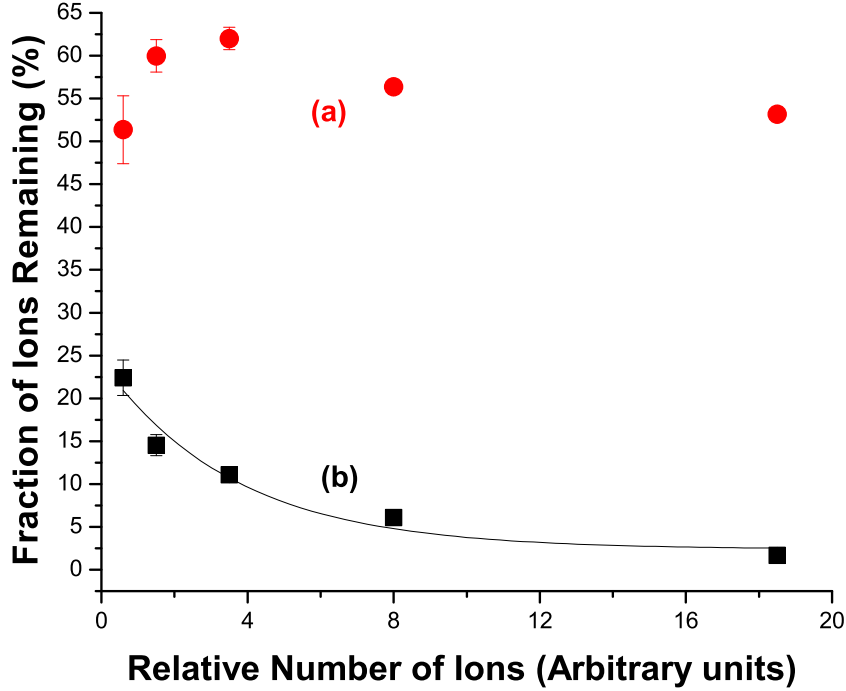}
\caption{(Color online). Experimental results showing evidence of the dependence of OREA trap loss on the relative initial number of ions. Different initial numbers of ions are loaded and extracted after a fixed $t_\mathrm{Trap}$ = 30 s for (a) $V_\mathrm{rad,\ce{Na2+}}$ = 0 V and  (b) $V_\mathrm{rad,\ce{Na2+}}$ = 2 V. The percent of ions lost from the trap for a given $t_\mathrm{Trap}$ is seen to increase exponentially with increasing initial number of ions. The fit in (b) is to a decreasing exponential. When not visible, error bars are smaller than the size of the points.}
\label{fig:Ion_number}
\end{figure}

The dependence of OREA on the number of trapped ions was also tested experimentally. This was tested by changing the initial number of \ce{Na+} ions loaded and extracting them after a fixed $t_\mathrm{Trap}$ (30 s). For each case $V_\mathrm{rad}$ was set at 2 V  or 0 V. The percent of initial \ce{Na+} ions remaining after 30 s, when $V_\mathrm{rad}$ = 0 V is relatively constant [as seen in Fig.~\ref{fig:Ion_number}(a)] regardless of the initial number of ions loaded into the trap. But when $V_\mathrm{rad}$ = 2 V, the percentage of initial ions remaining after 30 s is seen to decrease with increasing initial number of ions loaded. This result is consistent with our simulations. Experimentally, we observed an exponential dependence of trap loss (for a fixed trapping time of 30 s) on the number of ions trapped. This result shows that OREA is different from ion-ion rf heating, which has a linear dependence of the heating rate on the number of ions trapped \cite{Schuessler:05}.

Since $\omega_\mathrm{rad,\ce{Na+}} \approx 2.2 \omega_\mathrm{rad,\ce{Na2+}} $, we performed two experiments to see if the OREA on \ce{Na+} ions produced by the off resonant MSRQ  (set at $\omega_\mathrm{rad,\ce{Na2+}}$)  was due to discrete modes of instability of ion motion \cite{Drakoudis:2006}. Trapped ions' secular frequencies can be shifted by adding quadrupolar dc fields within the LPT trap region. This was accomplished by introducing dc fields on the rf segments of our LPT in the quadrupolar configuration. We performed a secular frequency scan, similar to the one described in Sec.~\ref{subsec:SIMResults}, for \ce{Na+} ions  with the added quadrupolar dc field. The results confirmed that  $\omega_\mathrm{rad,\ce{Na+}}$ was indeed shifted. Trap loss experiments performed with the additional quadrupolar dc field on the LPT showed that OREA increased even when the secular frequency of the trapped \ce{Na+} ions was shifted away from $2\omega_\mathrm{rad,\ce{Na2+}}$ (when MSRQ was set at $\omega_\mathrm{rad,\ce{Na2+}}$). Therefore, OREA is different from discrete ion motion instabilities. The addition of the dc quadrupolar field leads to different secular frequencies for the x and y motions of the ion ($\omega_1 \neq \omega_2$), which can cause additional trap loss even when the MSRQ field is removed.
 
\section{CONCLUSION}
\label{sec:Conclusions}

We have shown the effects of OREA due to MSRQ in a LPT filled with warm ions, via simulations and experiments. OREA can occur in any experiment that requires ejection or excitation of one ion species whose secular motion in an ion trap is off resonance with a different species of ion. Unwanted ions can be effectively ejected from a LPT by implementing MSRQ using either $V_\mathrm{rad}$ and/or $V_\mathrm{ax}$. Results of our on-resonance  simulations showed that, for our particular LPT geometry, using $V_\mathrm{rad}$  was the most efficient way to employ MSRQ. Our on-resonance experimental results confirmed the simulated prediction. We also simulated the dependence of OREA on the amplitude of the applied off-resonance field and the number of ions trapped. Experimentally we have observed that OREA trap loss time  has an exponential dependence  on the amplitude of the applied external fields ($V_\mathrm{rad}$ and $V_\mathrm{ax}$) and on the number of ions trapped, as confirmed by our simulations. 

\section{Acknowledgments}
\label{sec:Acknowledgments}

 We would like to acknowledge support from the NSF under Grants No. PHY–0855570 and  PHY1307874. One of us (FAN) would like to acknowledge the support from an ILIR grant.


\bibliographystyle{unsrt}
\bibliography{References}

\end{document}